\documentclass[pre,superscriptaddress,showpacs,nofootinbib,floatfix,preprint]{revtex4}
\usepackage{graphicx}
\usepackage{dcolumn}
\usepackage{bm}
\usepackage{epsfig}
\usepackage{booktabs,topcapt}
\usepackage{subfigure}
\begin{document}

\title{Hydration and anomalous solubility of the Bell-Lavis model as solvent}

\author{Marcia M. Szortyka\footnote[2]{e-mail - szortyka@gmail.com}}
\affiliation{Departamento de F\'{\i}sica, Universidade Federal 
de Santa Catarina, 
Caixa Postal 476, 88010-970, Florian\'opolis, SC, Brazil}

\author{Carlos E. Fiore\footnote[1]{e-mail - fiore@fisica.ufpr.br}}
\affiliation{Departamento de F\'isica, Universidade Federal do Paran\'a, 
Caixa Postal 19044, 81531 Curitiba, PR, Brazil}

\author{Marcia C. Barbosa\footnote[3]{e-mail - marcia.barbosa@ufrgs.br}}
\affiliation{Instituto de F\'isica, Universidade Federal do Rio Grande do
Sul, Caixa Postal 15051, 91501-970, Porto Alegre, RS, Brazil}

\author{Vera B. Henriques\footnote[4]{e-mail - vera@if.usp.br}}
\affiliation{Instituto de F\'{\i}sica, Universidade de S\~ao Paulo,
Caixa Postal 66318, 05315970, S\~ao Paulo, SP, Brazil}

\date{\today}
\begin{abstract}
We address the investigation of the solvation properties 
of the minimal orientational model for water, originally proposed by Bell and Lavis. The model presents two liquid phases separated by a critical line. The difference between the two phases is the presence of structure in the liquid of lower density, described through orientational order of particles. 
We have considered the effect of small inert solute on the solvent thermodynamic phases. Solute stabilizes the structure of solvent, by the organization of solvent particles around solute particles, at low temperatures. Thus, even at very high densities, the solution presents clusters of structured water particles surrounding solute inert particles, in a region in which pure solvent would be free of structure. Solute intercalates with solvent, a feature which has been suggested by experimental and atomistic simulation data. Examination of solute solubility has yielded a minimum in that property, which may be associated with the minimum found for noble gases. We have obtained a line of minimum solubility (TmS) across the phase diagram, accompanying the line of maximum in density (TMD). This coincidence is easily explained for non-interacting solute and it is in agreement with earlier results in the literature. We give a simple argument which suggests that interacting solute would dislocate TmS to higher temperatures. 
 \end{abstract}

\maketitle

\section{Introduction}

Biological molecules are functional only if organized spatially in very 
specific arrangements. This is the case for phospholipids in membranes, proteins 
soluble in water or membrane proteins, cholesterol or lipoproteins. One of the main 
ingredients behind spatial organization is solubility: globular proteins maintain 
their polar moieties on the exterior, in contact with water, while membrane protein 
must turn their polar parts inwards, avoiding contact with the hydrophobic bilayer core. 

Solubility depends on chemical structure, but varies with temperature. For simple 
substances, the behavior of solubility with temperature is dependent on 
miscibility, which describes the relative affinities of the molecules in 
solution \cite{Hi60}. 
The reasoning is simple. If we consider the solution phase in equilibrium 
with the gas phase, two situations exist. Consider X to be the solute in 
solvent Y. If Y and 
X 'prefer' mixing, which means that the energy of an YX pair is lower 
than the average energy of YY and XX pairs, for the solution energy to 
increase as temperature goes up, X must necessarily leave the solution, thus 
making solubility decrease. On the contrary, if Y and X prefer to phase 
separate, at low temperatures X will go preferentially to the gas phase. However, as 
temperature goes up, the solution energy increases while X dissolves in Y, making 
the solubility go up.

Solubility in water is different. Noble gases, for instance, present a temperature 
of minimum solubility in water at atmospheric pressure~\cite{Ba66}. Water presents 
in numerous thermodynamic and dynamic anomalies, and the minimum in solubility 
is one of them. The origin of the anomalies has been investigated theoretically 
both for statistical and atomistic models. However, a simple complete picture 
has not yet emerged.

The presence of a hydrogen-bond network was suggested in the '1930s by 
Bernal~\cite{Be13}, in order to explain the large mobility of H+ and OH- ions: the 
latter could only be explained if protons would jump between neighboring properly 
oriented molecules in liquid water. The idea of an extensive H bond network and a 
corresponding  water structure was probed with X-rays for many years, and 
the presence of the network was confirmed by more recent neutron scattering 
experiments, which pointed to an even more stable structure than previously 
believed~\cite{He02}. Hydrogen bonds are considered a key feature in
biochemistry~\cite{Pe97}.

The presence of an H-bond network could qualitatively explain the 
well-known maximum in density. The disordering of bonds allows density to increase
with temperature since the entropy of the bonds increases while translational entropy 
decreases, maintaining the necessary positive entropy balance.  The two entropic 
effects compete up to a temperature at which translational entropy wins over 
orientational entropy, taking density down, as in more 'usual' substances.

The dynamically connected molecules would also be able to explain the 
minimum in solubility. The contraction of the solvent, driven by decreasing
orientational entropy, excludes the solute. Thus a decreasing solubility is a 
consequence of an  increasing density of the solvent. In this case, the energy of 
the interactions enters  only 
either to favor the decreasing solubility, either to compete with it.

The study of statistical models capable of displaying properties typical of 
water has led, in the last years, to two basic models: (i) orientational 
models~\cite{Be94,Be70,Sa96,Ro96,Fr02a,Bu04a,Lo07,Al09}, which 
reflect the H-bonding property of water, and (ii) two-scale isotropic models, inspired 
on the low-temperature low-density property of water. Both models present 
several of the anomalous features of 
water~\cite{He70,Ja98,Wi01,Ca03,Ry03,Fo08,Fr01,Ol06a,Ba09,Ol10a,He05a,Fi09,Sz10a,Sz10b}. However, the second kind of 
model does not involve specific orientation of low energy pairs of particles: pair energy 
is controlled by distance, not by orientation. This poses a question of the 
relevance of the microscopic bonding in relation to the macroscopic properties.

In this study we propose to contribute to further investigation of 
the relation between the solvent structure and solubility.
The role of cavity formation in the explanation of hydrophobic interactions 
has been recognized by Pratt 
and Chandler~\cite{Pr77,Be96,Hu96}. They examine
the  difference between cavity formation 
in associating and simple liquids~\cite{Be96,Hu96}.
 A thorough investigation of noble gas solubility was
undertaken by Guillot and Guissani~\cite{Gu93} from  the 
point of view of atomistic models. Our approach is that of a 
minimal statistical model. We consider a two-dimensional lattice
model proposed originally by Bell and Lavis~\cite{Be70} and shown
by us~\cite{Fi09,Sz10b} to exhibit many anomalous properties in spite of the 
absence of liquid polymorphism. In this study we add non interacting
solute particles which occupy a single lattice site in order
to investigate the effect of solute on solvent properties as well as 
solute solubility. 

The remaining of the paper goes as follows. In sec. II the model without
and with solute 
is introduced, simulation details are presented in sec. III, the phase diagram
of the system with solute is shown in sec. IV, the solubility is 
analyzed in sec. V and conclusions are given in sec. VI.

\section{The Bell-Lavis model as solvent}

\begin{figure}[h!]
\begin{centering}
\includegraphics[scale=0.3,clip=true]{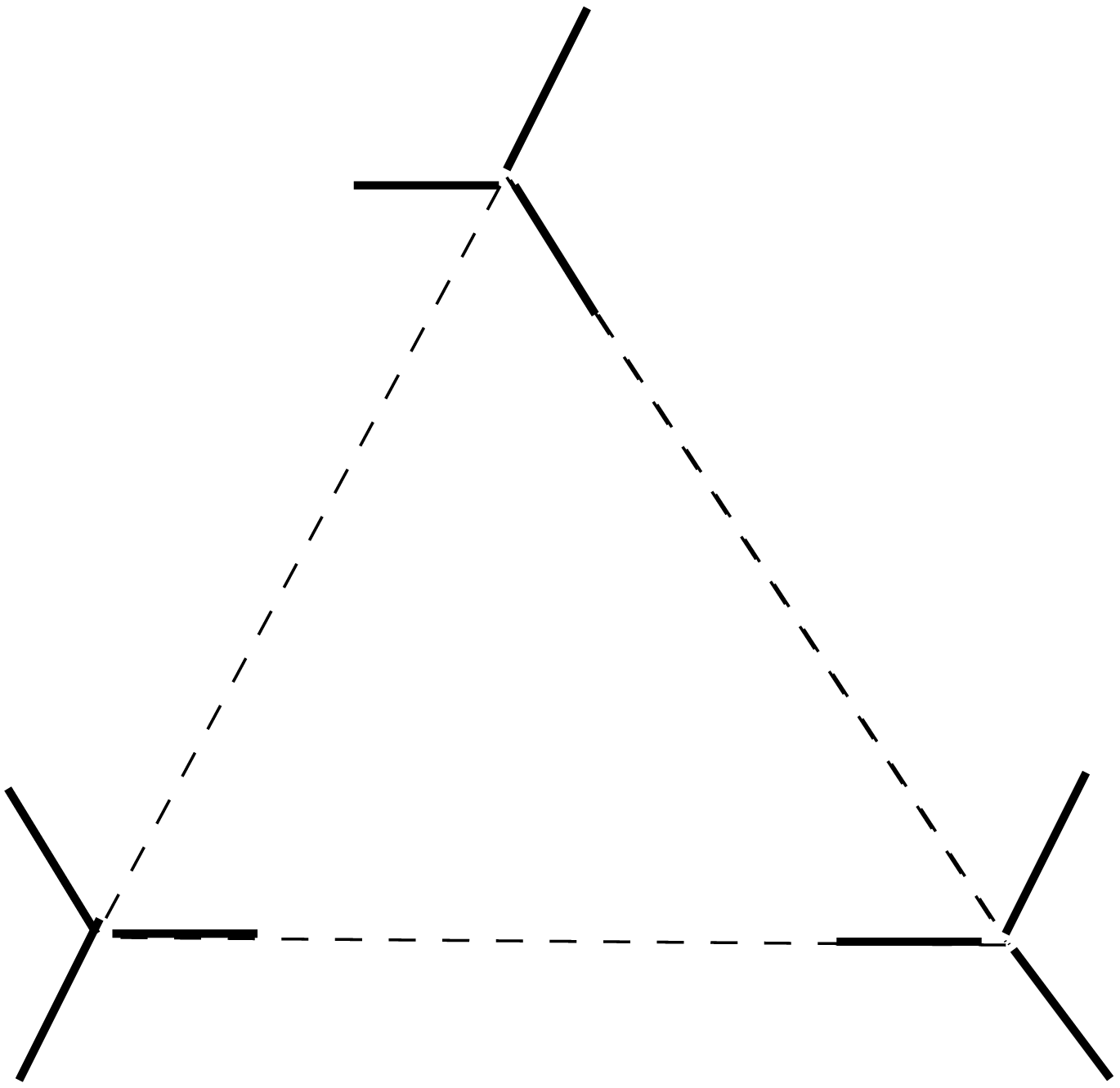} 
\hspace{2cm}
\includegraphics[scale=0.8,clip=true]{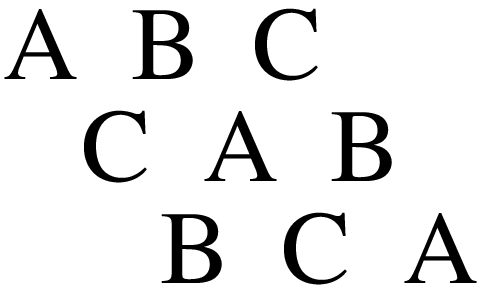} 
\par\end{centering}
\caption{(a) The arm variables in the triangular lattice. (b) The three
sub-lattices, A, B and C. }
\label{fig:model-sublattices} 
\end{figure}

The  Bell-Lavis (BL) model is defined on a triangular lattice where each 
site may be empty ($\sigma_i=0$) or occupied by an anisotropic water 
molecule ($\sigma_i=1$) \cite{Be70}. Each  particle has two 
orientational states, that may be described
in terms of six 'arm' variables
$\tau_{i}^{ij}$, with
 $\tau_{i}^{ij} = 1$ for the bonding state and $\tau_{i}^{ij} = 0$ for 
the inert arm state as illustrated in Fig.~\ref{fig:model-sublattices}(a).
A pair of adjacent molecules interacts via  van der Waals with energy  
$\epsilon_{vdw}$, as well as through
'hydrogen bonds' of energy $\epsilon_{hb}$, whenever bonding arms point 
to each other ($\tau_{i}^{ij}\tau_{j}^{ji} = 1$).
The model is defined by the following effective Hamiltonian in the 
grand-canonical 
ensemble
\begin{equation}                                                      
{\mathcal H} = -\sum_{<i,j>} \sigma_{i} \, \sigma_{j} \,      (\epsilon_{hb} \, \tau_{i}^{ij} \, \tau_{j}^{ji} +                     \epsilon_{vdw}) - \mu \sum_{i}^{} \sigma_{i},                  
\label{hambl}                                                                  \
\end{equation}
where $\epsilon_{vdw}$ and $\epsilon_{hb}$ are the van der Waals and 
hydrogen bond  interaction energies, respectively,and $\mu$ is the 
chemical potential. 

The model phase diagram features  depend on the ratio 
$\zeta=\epsilon_{vdw}/\epsilon_{hb}$ (see insets of 
Fig.~\ref{fig:mu.vs.t-0.02} that illustrates 
the reduced chemical potential versus
reduced temperature for two cases of bond
strength: weaker, $\zeta=1/4$,  and stronger,
$\zeta=1/10$). For  $\zeta<1/3$, besides the gas
phase, the model exhibits two 
liquid phases with different structure. At ${\bar T}=0$, 
coexistence between a gas and a structured liquid of 
low density (SL) as well as coexistence 
between the structured low density liquid and the
non-structured high-density-liquid (NSL) are 
present~\cite{Ba08}. However, for  finite temperatures, the 
transition between the two liquids becomes critical, as shown from 
detailed systematic analysis of simulational data~\cite{Fi09}. 
The two liquid phases do not coexist and the density varies 
continuously at the phase 
transition  as shown by susceptibility measurements on sub-lattices 
density fluctuations~\cite{Sz10b}. In order to stress the 
absence of a density gap we denominate the two liquid phases as 
structured (SL) and non-structured liquid (NSL), instead of adopting the 
usual LDL and HDL nomenclature. The difference between the two 
liquid phases lies in the orientational and 
translational order of the bonding particles. The SL phase presents a large 
population of particles in two of the three sub-lattices 
(see Fig.~\ref{fig:model-sublattices}(b)) associated 
to a large bonding network, whereas in the NSL  the density is close to 1 
and orientational order is lost. The increase in the temperature
and the increase in the chemical potential favor 
the NSL phase.  In the case of the stronger
hydrogen bonds ($\zeta=1/10$) the SL is favored and the 
transition occurs for higher chemical potentials.

A line of temperatures of maximum density (TMD) lies near the critical 
line separating  the two  liquid phases. Its pressure and temperature
location is not very sensitive to the strength of the 
hydrogen bonds, in both in the $\zeta=1/10$ and in the $\zeta=1/4$ cases
at low chemical potentials is located in the SL phase while for
high chemical potentials is located at the critical line .
In this work we 
 have added inert apolar solutes to the BL model. The new particles 
occupy empty sites and thus interact only via excluded volume with the 
BL solvent particles. Our purpose is the investigation of the effect of the 
apolar solute upon the TMD and the regions of stability of the low and 
high density phases.
Here we address the following questions.
 What would be the effect of adding solute to 
the structured liquid? Under what 
circumstances does phase separation occur? Is there a 
solubility minimum? In the 
latter case, can we establish a relation between the 
density and the solubility 
anomalies?

\section{Monte Carlo simulations}

We have investigated the properties of our model solution through 
Monte Carlo simulations, in a mixed ensemble, of fixed chemical potential 
for solvent and constant density for the apolar solute, under periodic 
boundary conditions.

The model solvent microscopic configurations were generated through 
randomly selected exclusion, insertion or rotation of water 
particles, whereas  solute movements were based on solvent-solute and  
hole-solute exchanges.  Acceptance rates are those of 
the usual Metropolis algorithm: transitions between two configurations are 
accepted according to
the Metropolis prescription $\min\{1,\exp(-\beta \Delta {\cal H})\}$, where
$\Delta {\cal H}$ is the effective energy difference between the two
states.
Our simulations were carried out for lattice
sizes ranging from $L=30$ to $L=60$. Results
shown here are for $L=30$. All the thermodynamic 
quantities are expressed in reduced units of $\epsilon_{hb}$ 
and lattice distance.

\section{Solvent Phase Diagram in the presence of inert solute}

We have investigated how the chemical potential 
versus temperature phase diagram changes by 
the addition of an inert  solute. We study this employing
two   solute concentrations, $2\%$ and $10\%$. 

\begin{figure}[h!]
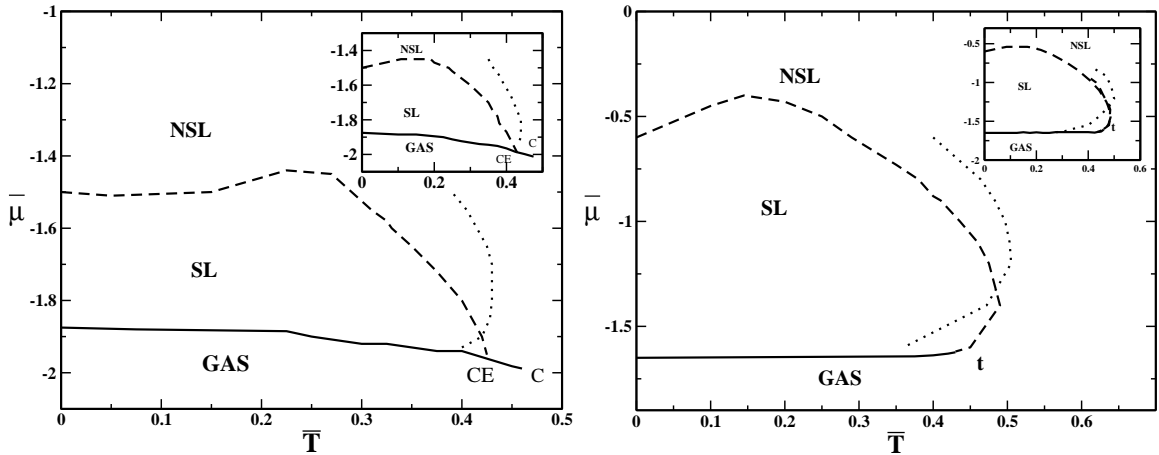

\begin{centering}
\includegraphics[scale=0.3,clip=true]{diagrama_zeta_1-4_c_2_l30.eps} 
\includegraphics[scale=0.3,clip=true]{diagrama_zeta_1-10_c_2_l30.eps} 
\par\end{centering}
\caption{Solvent chemical potential $\bar{\mu}$ vs. reduced
temperature $\bar{T}$  phase diagram for model 
solution at solute concentration $2\%$ 
and $\zeta=1/4$ (left) and $\zeta=1/10$ (right). 
Solid, dashed and dotted lines 
correspond to first-order, second-order phase transitions
and the  TMD, respectively. The symbols C, CE and $t$ denote
critical, critical-ending and tricritical points, respectively.
The inset displays the corresponding phase diagram 
for pure solvent, for comparison.}
\label{fig:mu.vs.t-0.02} 
\end{figure}

The reduced chemical potential versus reduced
temperature phase diagrams for both
weak and strong bonds,  $\zeta=1/4$ and $\zeta=1/10$, and  concentration 
of  $2\%$ of solute,  are shown in Fig. \ref{fig:mu.vs.t-0.02}. At this 
small concentration of solute, the phase diagram suffers small 
quantitative changes: the structures phase SL extends to slightly 
higher chemical potential, while, in the case of weaker 
bonds, $\zeta=1/4$, the TMD line moves into the SL phase at 
low temperature. Thus solute 
stabilizes the SL to higher chemical potential, which suggests a 
reinforcement of hydrogen bonding. It also brings down the 
temperature of maximum density in the case of weaker bonds, turning
 TMD behavior similar to the case of stronger bonds - again, solute 
seems to "strengthen" bonds. 


\begin{figure}[h!]
\begin{centering}
\includegraphics[scale=0.7,clip=true]{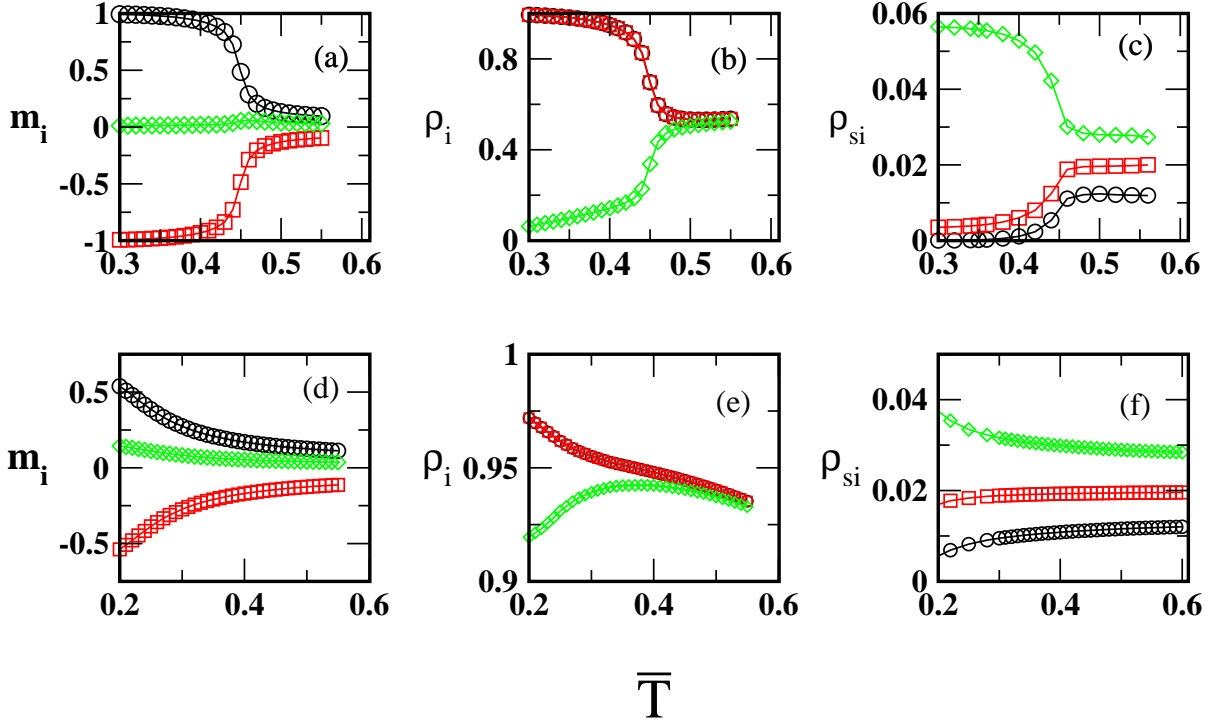} 
\par\end{centering}
\caption{Sub-lattices distributions for solvent and solute properties 
for $2\%$ solute concentration and $\zeta=1/10$. From right 
to left, we have solvent particle 
orientation $m_i$ , solvent particle density $\rho_i$ 
and solute density $\rho_{si}$ versus reduced
temperature $\overline {T}$ . Top graphs are for lower reduced
chemical potential $\bar{\mu}= -1.6$, bottom graphs are for higher reduced
chemical potential $\bar{\mu}= -0.4$. Colors identify sub-lattices.}
\label{fig:rho.vs.t-0.02} 
\end{figure}

Fig.~\ref{fig:rho.vs.t-0.02} illustrates features 
of the solution structure in the 
case of strong bonds, $\zeta=1/10$ and concentration
of $2\%$ of solute. The triangular lattice is subdivided into 
three sub-lattice as illustrated in Fig.~\ref{fig:model-sublattices}(b).
The orientation and density of solvent particles, as well as, the
 density of solute are computed on each sublattice. 

The first set of data 
at Fig.~\ref{fig:rho.vs.t-0.02}(a)-(c) illustrate the orientation of 
the solvent molecules, the density of solvent and the density of the solute versus 
temperature for $\bar{\mu}=-1.6$. The graphs show the 
transition between the NSL to the SL phase by
decreasing the temperature. It can be seen that in the SL 
phase ($\bar{\mu}=-1.6$)  solvent occupies mainly two of the sub-lattice 
(with $\rho_{solvent}$ nearly 1 at $\bar{T}=0.3$), with complementary 
orientations ($m=+1$ and $m=-1$), indicating strong bonding. Both quantities 
vary abruptly at the transition to the NSL phase, around $\bar{T}=0.45$, with 
homogeneous occupation and orientation of molecules on the three 
sub-lattice. As for solute, at the lower 
temperature, $\bar{T}=0.3$, occupation of the empty sublattice is 
preferential ($\rho_{solute} \approx 2\%$), while 
the other sub-lattice are nearly empty. As solvent disorders on 
sub-lattice, around $\bar{T}=0.45$, solute densities vary continuously 
towards homogeneous occupations of the three sub-lattice.

 Fig.~\ref{fig:rho.vs.t-0.02}(e)-(f) illustrate
the same data as before but for reduced chemical potential $\bar{\mu}=-0.4$.
In this case, no transition is observed. The system is in the NSL 
phase even at low temperature, and sub-lattice solvent orientation and 
solute density vary continuously towards homogeneous distribution 
on sub-lattice. Solvent density still carries the signature of the 
ordered phase, transitioning smoothly to disorder in a sigmoidal fashion.

\begin{figure}[h!]
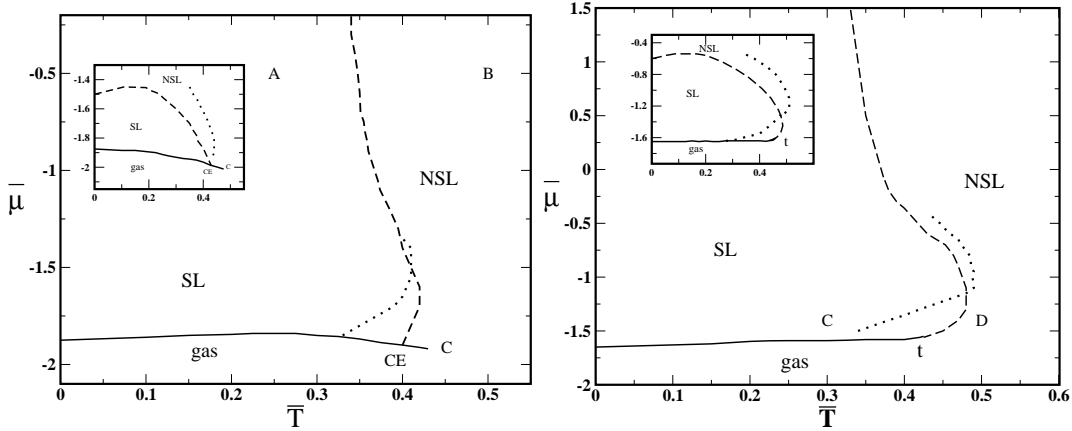

\begin{centering}
\includegraphics[scale=0.3,clip=true]{diagrama_zeta_1-4_c_10_l30.eps}
\includegraphics[scale=0.3,clip=true]{diagrama_zeta_1-10_c_10_l30.eps} 
\par\end{centering}
\caption{Solvent reduced chemical potential $\bar{\mu}$ versus
reduced temperature $\bar{T}$  phase diagram for model solution 
at solute concentration $10\%$ 
for $\zeta=1/4$ and $\zeta=1/10$.
Solid, dashed and dotted lines 
correspond to first-order, second-order phase transitions
and the  TMD, respectively. The symbols $CE$, $C$ and  $t$ denote
critical-ending, critical and tricritical points, respectively.
The insets displays the corresponding phase diagrams for 
pure solvents for comparison. }
\label{fig:mu.vs.t-0.10} 
\end{figure}

Figure~\ref{fig:mu.vs.t-0.10}  displays the reduced
chemical potential versus reduced temperature 
phase diagrams for  $10\%$ concentration of the solute,  for both values 
of hydrogen bond strength $\zeta=1/4$ (left) and  $\zeta=1/10$ (right). In this case 
substantial change in the phase diagrams can be seen. The low temperature SL to NSL phase 
transition seen as one increases chemical potential for pure solvent 
is destroyed by the presence of solute. Instead, the transition may be 
reached only from temperature variations, and the SL phase extends to very 
high chemical potentials. The TMD line moves nearer to the critical SL-NSL 
line and crosses into the SL phase.

\begin{figure}[h!]
\begin{centering}
\includegraphics[scale=0.6,clip=true]{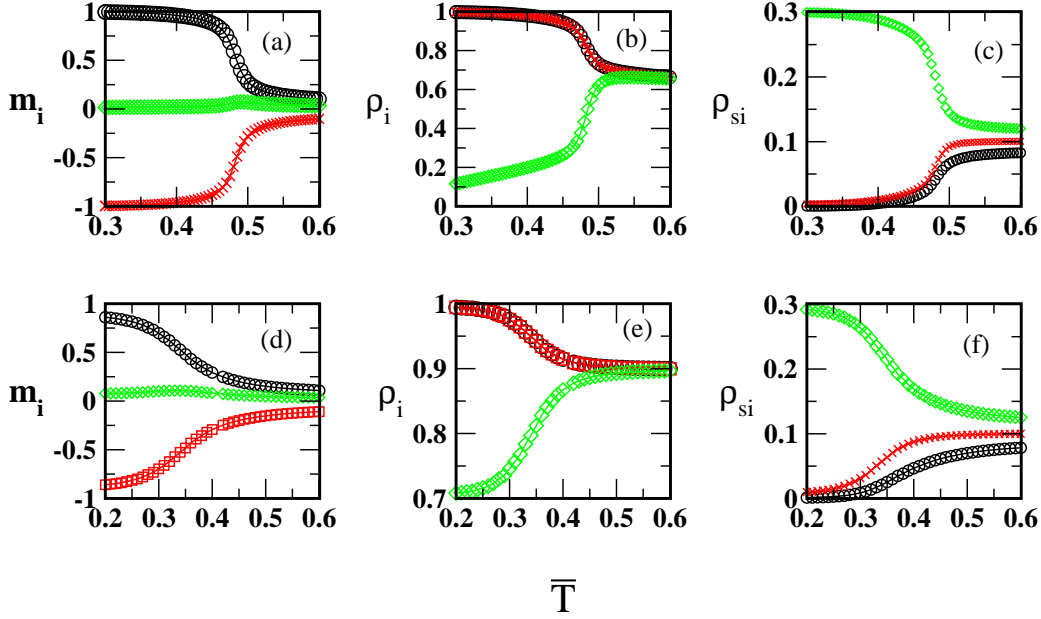} 
\par\end{centering}
\caption{Solution structure in the new phase. Sub-Lattice 
solvent particle orientation $m_i$, solvent density 
$\rho_i$,  and  solute density $\rho_{si}$ vs. 
temperature  for different chemical potentials 
 for  $10\%$ of solute concentration and $\zeta=1/10$. The top 
and bottom graphs correspond to solvent chemical potentials $\mu=-1.30$ and $\mu=4.0$. }
\label{fig:rho.vs.t-0.10} 
\end{figure}

In Fig.~\ref{fig:rho.vs.t-0.10} we investigate the solvent orientation 
and density, as well as, the density of solute in each sublattice  
in different regions of the phase diagram at $10\%$ solute concentration 
for bond strength $\zeta=1/10$. For both low, $\bar{\mu}=-1.3$, and high, 
$\bar{\mu}=4.0$,  reduced chemical 
potential, solute orders together with solvent at low temperatures. As 
can be seen from the color identification of lattices, solute goes 
into the empty lattice while solvent particles orient properly on 
two sub-lattice in order to connect through bonds. 

For both reduced chemical potentials an abrupt variation of the solvent 
density, of the solute density and of the solvent orientation occur
simultaneously, near $\bar{T}=0.45$ for $\bar{\mu}=-1.3$ and
near $\bar{T}=0.3$ for $\bar{\mu}=4.0$ at the SL-NSL 
transition line. 

However, despite of the qualitative similar behavior
there are quantitative important
differences between the two regions of chemical potential.
 For the lower chemical potential, $\bar{\mu}=-1.3$, solvent 
behavior is similar to that of pure solvent. The orientationally 
ordered solvent particles occupy mainly two of the sub-lattice, while 
the third sublattice remains nearly free of solvent. At $\bar{T}=0.3$ 
nearly $60\%$ 
of that sub-lattice stands vacant, while solute particles occupy $30\%$ 
of the sites, leaving the other two sub-lattice free of solute.

At high chemical potential, $\bar{\mu}=4.0$, while solute maintains 
the $30\%$ occupancy of one of the sub-lattice at low temperature, the 
solvent particles fill up the rest of the sublattice sites, reaching 
$70\%$ occupancy of that sublattice.

The new behavior induced by the presence of solute is better understood
by comparing Fig.~\ref{fig:rho.vs.t-0.02}(e) and 
Fig.~\ref{fig:rho.vs.t-0.10}(e).
Differently from the $2\%$ solute concentration, for the  $10\%$  
concentration case the filling up of the lattice yields only partial 
rupture of hydrogen-bonding. Maintenance of the hydrogen 
bond network at such high density seems to be a result of
 the structuring effect of solute.

\begin{figure}[h!]
\begin{centering}
\includegraphics[scale=0.2,clip=true]{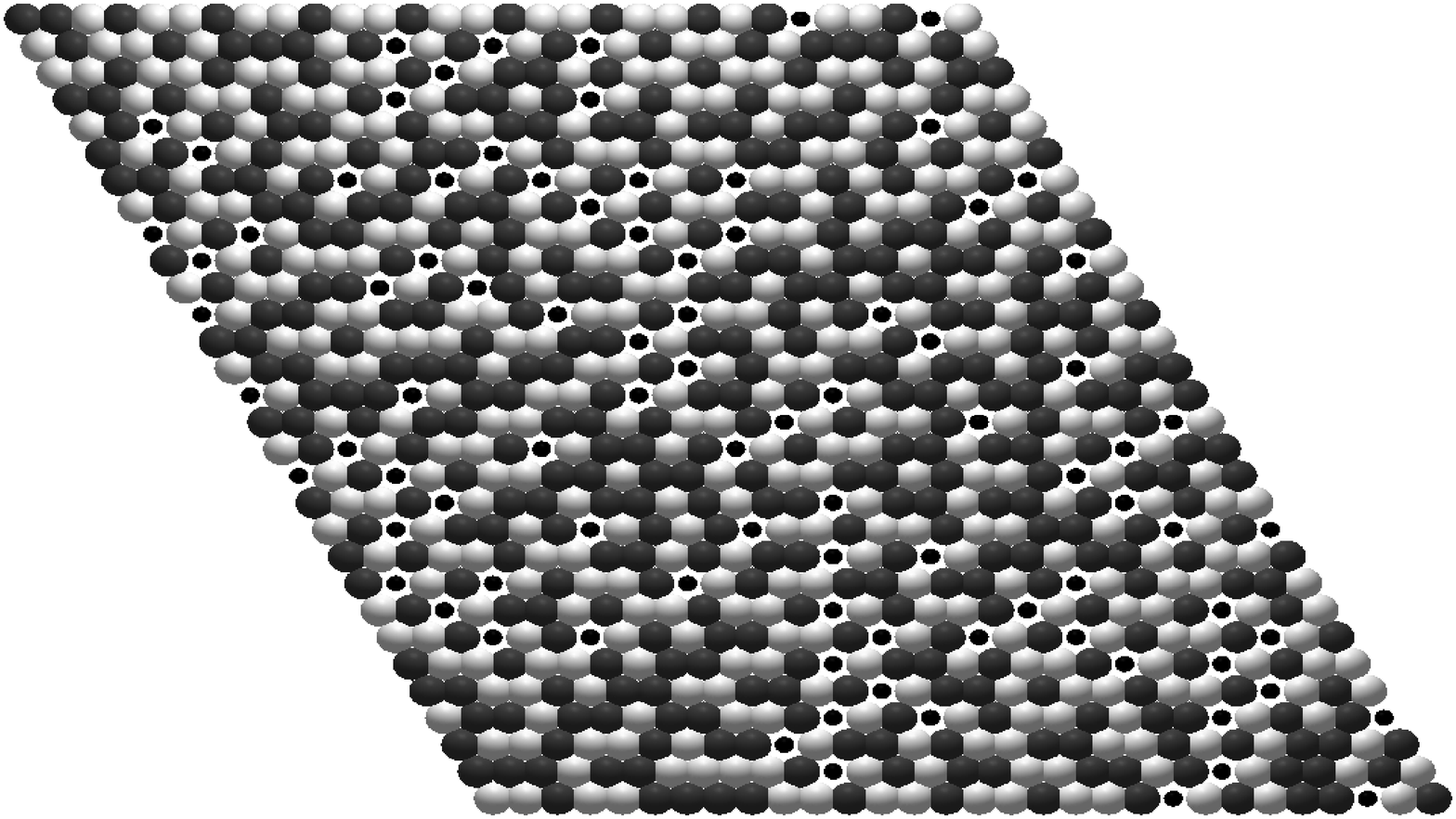} 
\includegraphics[scale=0.2,clip=true]{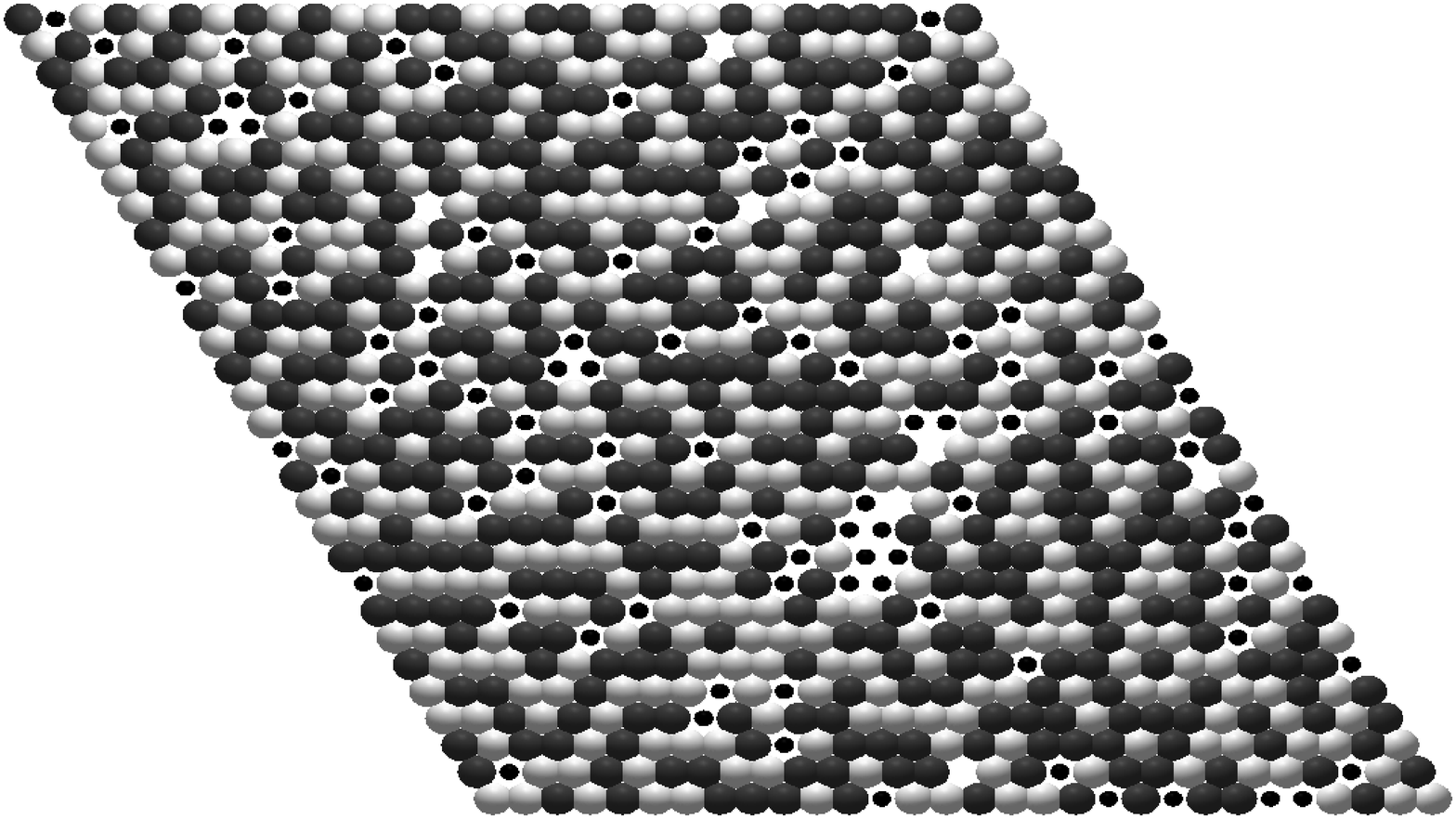} 
\par\end{centering}
\caption{Snapshots of model system for concentration of $10\%$ and $\zeta=1/4$
at high chemical potential, $\bar{\mu}=-0.5$, in A ($\bar{T}=0.25$, left) and B ($\bar{T}=0.5$, right)
 points in Fig.~\ref{fig:mu.vs.t-0.10}. 
Black and Grey circles are two orientations of solvent particles, black dots represent solute particles.   }
\label{fig:snapshots2} 
\end{figure}

\begin{figure}[h!]
\begin{centering}
\includegraphics[scale=0.2,clip=true]{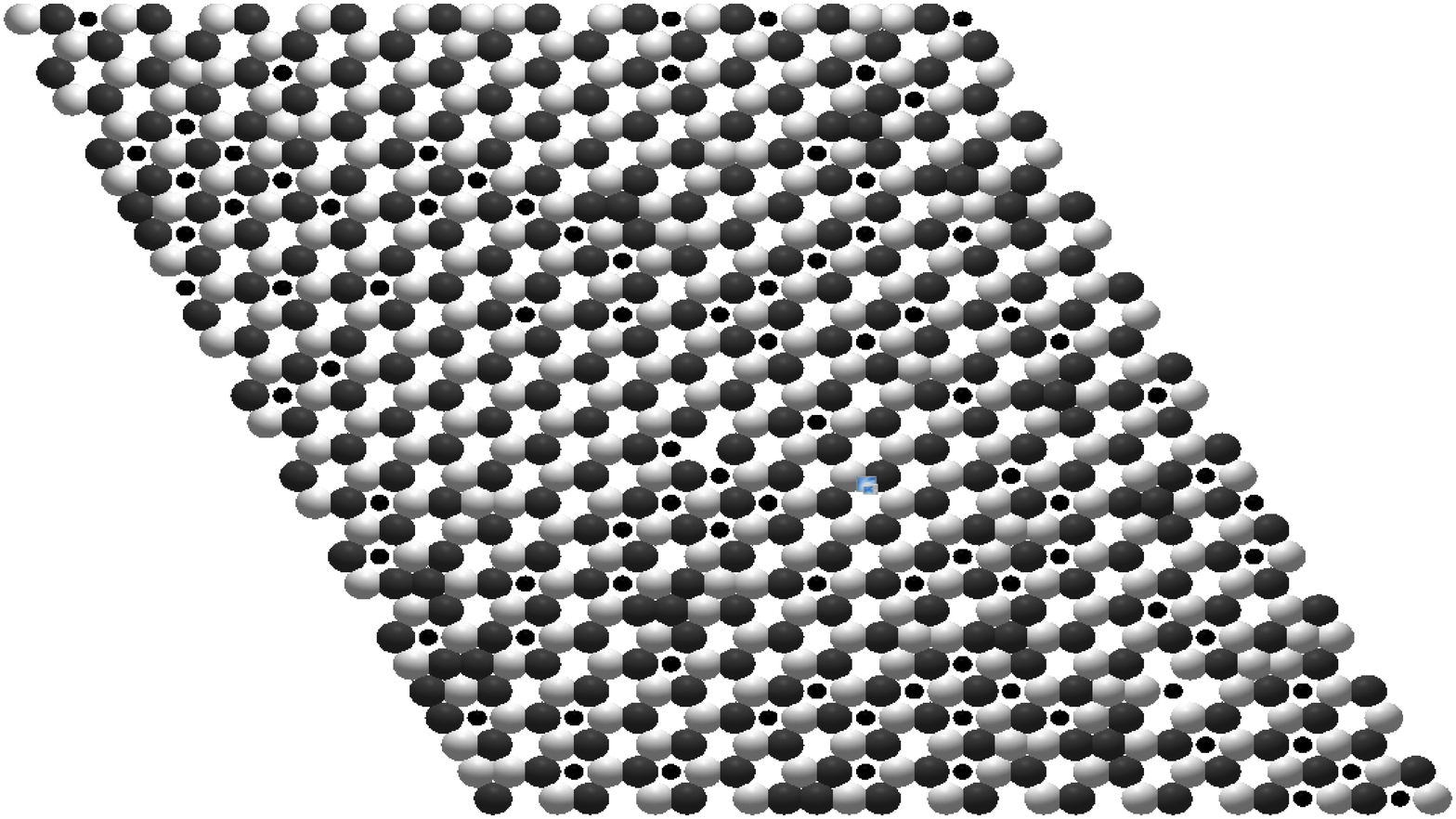} 
\includegraphics[scale=0.2,clip=true]{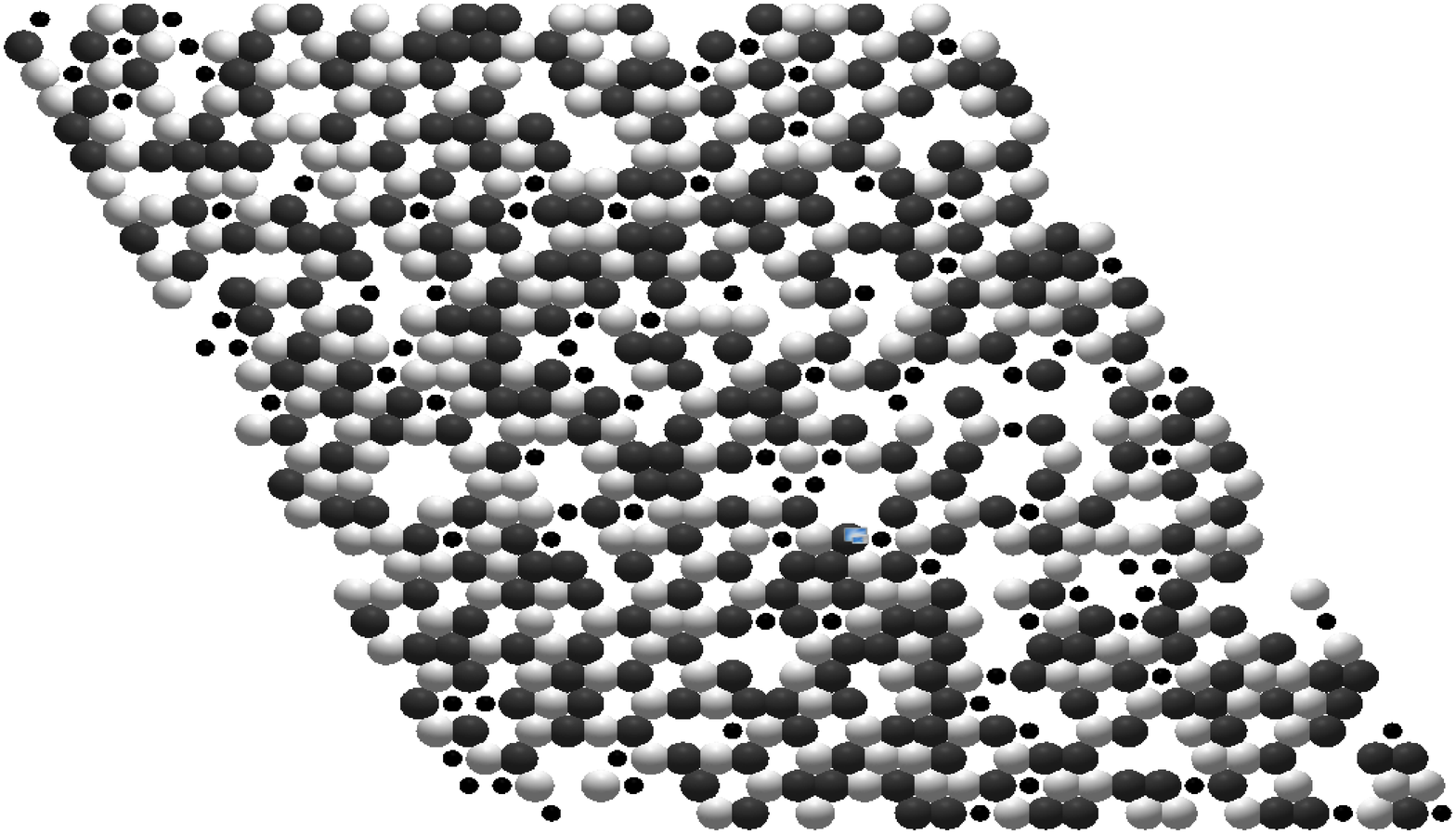} 
\par\end{centering}
\caption{Snapshots of model system for concentration of $10\%$ and $\zeta=1/10$ at 
low chemical potential, $\bar{\mu}=-1.4$, in C ($\bar{T}=0.3$, left) and 
D ($\bar{T}=0.5$, right) points in Fig.~\ref{fig:mu.vs.t-0.10}. Black and gray circles are two orientations of 
solvent particles, black dots represent solute particles.   }
\label{fig:snapshots} 
\end{figure}

Inspection of typical configurations in different regions of the phase 
diagrams are quite useful at this point. Fig.~\ref{fig:snapshots} 
and Fig.~\ref{fig:snapshots2} display
snapshots of the model system at different points 
(indicated by letters in Fig.~\ref{fig:mu.vs.t-0.10}) 
 in the reduced chemical potential versus reduced temperature  phase diagrams 
with solute concentration $10\%$.

For $\zeta = 1/4$ (Fig.~\ref{fig:snapshots2}), at $\bar{\mu} = -0.50$ 
and $\bar{T}=0.25$ (inside SL phase, point A in Fig.~\ref{fig:mu.vs.t-0.10} ), 
lattice is filled up, but patches of 
structured liquid can be seen with solute localizing only in sites 
which contribute to organize the hydrogen bond network. As the SL-NSL 
line is crossed and for $\bar{T}=0.50$ (point D in Fig.~\ref{fig:mu.vs.t-0.10}), 
a few isolated solute particles are 
surrounded by water particle structure, while most solute particles are 
clustered  in vacant regions.

For $\zeta = 1/10$ (Fig.~\ref{fig:snapshots}),
at $\bar{\mu} = -1.40$ and $\bar{T} = 0.30<\bar{T}_{TMD}$ (point C in Fig.~\ref{fig:mu.vs.t-0.10}),
 a fully bonded network of solvent particles is accompanied by solute particles located in the 
empty sublattice.  This gives rise to  apparently linear aggregates 
intercalated by solvent. At a temperature higher than 
the TMD, $\bar{T}=0.50$ (point D in Fig.~\ref{fig:mu.vs.t-0.10}), some bonding 
of the solvent particles in hexagons are  still seen, with 
intercalated solute.However, the  system is much less dense, and  
solute particles also  
localize in large vacant regions.

\section{Model solubility}


The Ostwald solubility $\Sigma$  is defined  as the ratio between 
solute densities $\rho_X$ in the two coexisting   phases:
\begin{equation}                                                             
\Sigma=\frac{\rho_{X}^{I}}{\rho_{X}^{II}}.                                    
\end{equation}
The two coexisting phases, $I$ and $II$, might either be  a gas 
phase $II$ that coexists with a homogeneous liquid phase 
$I$~\cite{Gu93} or two liquid phases $I$ and $II$,
 of different relative densities, the first poor in solute $X$, the 
other reach in $X$~\cite{Ba91}. 

In the case of liquid-liquid phase separation, the form of the 
temperature-density coexistence curve, at fixed pressure, is 
indicative of solubility behavior. If the density gap decreases
 as temperature is increased, solubility increases with 
temperature. However, for reentrant coexistence curves, for 
which the density gap increases as temperature is increased, solubility 
decreases as temperature is raised.

On the other hand, minimal statistical models show that for dense lattice
gas solutions 
with isotropic van der Waals-like interactions solubility behaves 
univocally with temperature. Coexistence densities between an ideal 
gas mixture and a lattice dense solution, for substances Y and X, are 
obtained from the equality of the corresponding chemical potentials. Consider
 interaction constants $w_{YY}$, $w_{XX}$ and $w_{YX}$ between pairs 
YY, XX and YX. For solute X in the gas phase given by the  
dimensionless solute X density $\rho_{X}^{gas}$, we might write
\begin{equation}
\mu_{X}^{gas}=k_{B}T\ln \rho_{X}^{gas},
\label{eq:mugas}
\end{equation}
whereas for solute X in the dense lattice solution, we have
\begin{equation}
\mu_{X}^{solution}=\frac{ w_{XX}}{2}-\frac{w}{2}(1-x_X)^2
+k_{B}T\ln x_{X}^{solution},
\label{eq:musolution}
\end{equation}
where $w= w_{YY}+w_{XX}-2w_{YX}$ and $x_X$ is the solution concentration given in mole fraction. 
Equating Eq. (\ref{eq:mugas}) to Eq. (\ref{eq:musolution}) yields
\begin{equation}                                                             
\frac{x_X^{solution}}{\rho_X^{gas}}= e^{-\frac{\beta w_{XX}}{2}}
e^{\frac{\beta w}{2}(1-x_X^{solution})^2}\; .
\label{eq:rhoXsolution}
\end{equation}
A slightly different definition of solubility, proportional to the inverse of Henry's constant, is given by
\begin{equation}                                                             
\Sigma'=\frac{x_{X}^{solution}}{\rho_{X}^{gas}/\rho_{X}^{0}},  
 \label{eq:rhoXsolution'}      
\end{equation}
where $\rho_X^0$ is gas density for pure liquid X. Comparing
Eq. (\ref{eq:rhoXsolution}) with Eq. (\ref{eq:rhoXsolution'}) gives
\begin{equation}                                                             
\Sigma'=e^{\frac{\beta w}{2}(1-x_X^{solution})^2}.
\end{equation}
 Thus for poorly miscible solutions, with $w<0$, which 
phase  separate at low temperatures, solubility increases with 
temperature, since $d\Sigma'/dT \propto  -w$. On the other hand, if the 
two liquids are miscible, when 
$w>0$, solubility decreases as temperature is raised. In either 
case, solubility displays monotonic behavior with temperature. 

The solubility behavior of the dense lattice model is the result of a competition between entropy of mixture and an isotropic interaction potential. relies on positional entropy
The model misses the role of density, an essential feature of water. 

\begin{figure}[h!]
\begin{centering}
\includegraphics[scale=0.5,clip=true]{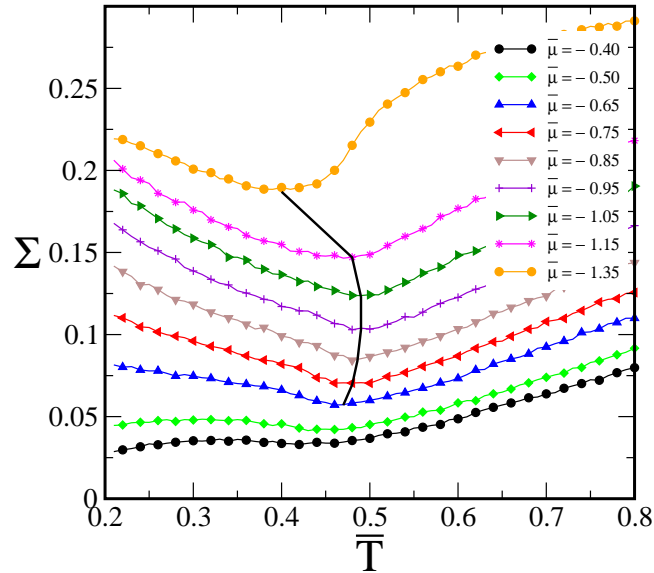} 
\par\end{centering}
\caption{Ostwald coefficient $\Sigma$ versus $\overline {T}$ 
for different $\overline {\mu}$ for $\zeta=1/10$ and $10\%$. The 
black line corresponds the temperature for which the density
presents a  maximum. }
\label{fig:sigma.vs.t} 
\end{figure}

How does the introduction of asymmetry in the interaction 
potential, accompanied by orientational entropy, change this picture? 
In order to answer to this question we have measured the solubility of 
our model inert solute as a function of temperature for different 
fixed chemical potential of solvent, by assuming coexistence of a 
gas (phase $Y$) and a homogeneous solution phase ($X$). The gas 
phase was assumed ideal, thus
\begin{equation}
\mu_{X}^{gas}=-k_{B}T\ln \rho_{X}^{gas}.
\end{equation}
For the solution phase, the chemical potential of solute was 
calculated from simulation data through Widom's insertion 
method~\cite{Wi63}. In our semi-grand canonical 
ensemble the semi-grand potential $\psi=\psi(T,V,N_{X},\mu)$ 
depends on $T$, $V$, $N_X$ and on the solvent chemical potential 
$\mu$. In the thermodynamic limit, the solute chemical potential
$\mu_{X}^{solution}=-(\frac{\partial \psi}{\partial N_X})$ can be 
approximated
by the difference $\psi(T,V,N_X+1,\mu)-\psi(T,V,N_{X},\mu)$ 
and we have
\begin{equation}
\mu_{X}^{solution}=-k_{B}T\ln \frac{\Xi(T,V,\mu,N_{X+1})}{\Xi(T,V,\mu,N_{X})},
\end{equation}
which relates average values in two different ensembles of $N_{X}$ 
and $N_{X+1}$ particles. However, the numerator can be interpreted in 
terms of an average in the ensemble of $N_{X}$ solute particles. Thus we have 
\begin{equation}
\mu_{X}^{solution}=-k_{B}T \ln \left( \frac{1}{\langle\rho_{X}^{solution}\rangle}\right)
\langle e^{-\beta \Delta u}\rangle_{T,V,N_X,\mu},
\label{eq5s}
\end{equation}
where $\Delta u$ is the additional energy due to insertion 
of solute molecule to a system 
of solute concentration $\rho_{X}^{solution}= N_X/V$. 
Finally, by equating $\mu_{X}^{solution}$ and $\mu_{X}^{gas}$,
we obtain for the solubility
\begin{equation}                                                             
\Sigma=\frac{\rho_{X}^{solution}}{\rho_{X}^{gas}}=\langle   
e^{-\beta \Delta u}\rangle_{T,V,N_X,\mu}.
\label{eq:sol}                                    
\end{equation}
In the Fig.~\ref{fig:sigma.vs.t} we display our data 
for solubility $\Sigma$ versus temperature for bond
strength $\zeta=1/10$. As can be seen, a 
minimum is present for different chemical potential of solvent. 

The temperature of minimum solubility (TmS) coincides entirely 
with the temperature of maximum density (TMD), in 
the $\bar{\mu}$  vs. $\bar{T}$ plane. This is to be expected for inert solutes. Inspection 
of  Eq. (\ref{eq:sol}) for inert solutes yields 
$\exp \{-\beta \Delta u\}=1$ for insertion 
into empty sites and $0$, otherwise. Thus solubility can be directly 
related to the overall liquid density $(N_X+N^{solvent})/V$. Thus
\begin{equation}                                                             
\Sigma=\langle e^{-\beta \Delta u}\rangle_{T,V,N_X,\mu}=      
1-\rho^{solvent}-\rho_{X}^{solute} ,
\end{equation}
and for fixed solute density
\begin{equation}                                                             
\frac{d\Sigma}{dT}= - \frac{d\rho^{solvent}}{dT}, 
\end{equation}
and therefore the TMD is accompanied by the TmS.
Is the coincidence between TmS and TMD restricted to inert solutes? 

It is tempting to extend our analysis of Eq. (\ref{eq:sol}) to interacting solutes. A first 
simplest approach to the question would be to investigate the energetic 
effect on solubility through the following approximation

\begin{equation}                                                             
\Sigma=\langle e^{-\beta \Delta u}\rangle_{T,V,N_X,\mu}\approx   
(1-\rho^{solvent}-\rho_{X}^{solute} ) e^{-\beta <\Delta u>},
\end{equation}
thus
\begin{equation}                                                             
\frac{d\Sigma}{dT}\approx \left[- \frac{d\rho^{solvent}}{dT}+(1-\rho^{solvent}-\rho_X^{solute})\left(
\frac{<\Delta u>}{k_BT^2}-\beta \frac{d <\Delta u>}{dT}\right)\right] e^{-\beta < \Delta u>}.
\end{equation}
Since $<\Delta u>$ is necessarily negative and $\frac {d<\Delta u>}{dT}$ is necessarily positive, this result implies that the minimum in 
solubility should occur at a temperature higher than TMD. This is in accordance 
with data on solubility of gases.

\section{Conclusions}

In this paper we have considered the investigation of the thermodynamic phases 
and of solubility of the Bell-Lavis (BL) water model 
in the presence of small inert solute. The Bell-Lavis two-dimensional orientational 
model presents a density anomaly and two liquid phases of different 
structure\footnote{a modified form of the BL model has been investigated as to solvation entropy and enthalpy properties~\cite{Bu05}}.

We have considered two fixed concentrations of solute, respectively $2\%$ and $10\%$. In 
both cases, but more evidently for $10\%$ solute, the presence of solute 'strengthens' 
the hydrogen bonds. Inspection and comparison of phase diagrams show that the 
structured phase is stabilized to higher temperatures, at fixed chemical potential, and 
to much higher chemical potentials. For the higher concentration of solute, the transition 
between the structured and the unstructured phases as chemical potential is varied 
disappears. Examination of solvent structure shows that the presence of solute 
nucleates patches of hydrogen-bonded solvent particles. Solute intercalates with 
properly oriented solvent. Both features, increment of water structure and 
solvent-separated solute states, have been reported from experiments and 
atomistic models~\cite{La93}.

Solubility of our small inert solutes presents a minimum (TmS), which coincides 
with the maximum solvent density (TMD), as expected~\cite{Pr77}. For interacting solute, a simple 
argument leads us to expect TmS to occur at higher temperatures for the BL solvent model.

Investigation of the latter point, as well as the effect of solute size are the 
subject of ongoing work.

\section*{ACKNOWLEDGMENTS}

We thank for financial support the Brazilian science agencies
CNPq and Capes. This work is partially supported by CNPq, INCT-FCx.

\end{document}